\documentclass[aps,pra,twocolumn]{revtex4}
\newcommand {\be}{\begin{equation}}
\newcommand {\ee}{\end{equation}}
\newcommand {\bey}{\begin{eqnarray}}
\newcommand {\eey}{\end{eqnarray}}

\usepackage{epsfig}
\usepackage{amsfonts}
\usepackage{amssymb}
\usepackage{amsmath}
\usepackage{enumerate}
\usepackage{amsthm}
\usepackage{pgfplots}
\usepackage{tikz}

\usepackage{hyperref}

\newtheorem{definition}{Definition}
\newtheorem{theorem}{Theorem}
\theoremstyle{definition}
\newtheorem{problem}{Problem}
\newtheorem{lemma}{Lemma}
\newtheorem{corollary}{Corollary}
\newtheorem{algorithm}{Algorithm}

\begin{document}

\title{Can non-local correlations be discriminated in polynomial time?}
\author{Alberto Montina, Stefan Wolf}
\affiliation{Facolt\`a di Informatica, 
Universit\`a della Svizzera italiana, 6900 Lugano, Switzerland}
\date{\today}

\begin{abstract}
In view of the importance of quantum non-locality in 
cryptography, quantum computation and communication complexity, 
it is crucial to decide whether a given correlation exhibits 
non-locality or not. In the light of a theorem by Pitowski,
it is generally believed that this problem is computationally 
intractable. In this paper, we first prove that the Euclidean 
distance of given correlations from the local polytope can be 
computed in polynomial time with arbitrary fixed error, granted 
the access to a certain oracle. Namely, given a fixed error,
we derive two upper bounds on the running time. The first
bound is linear in the number of measurements.
The second bound scales as the number of measurements to the sixth 
power. The former is dominant only for a very high number of
measurements and is never saturated in the performed numerical
tests.
We then introduce a simple algorithm for simulating the oracle. 
In all the considered numerical tests, the simulation of the oracle
contributes with a multiplicative factor to the overall 
running time and, thus, 
does not affect the sixth-power law of the oracle-assisted 
algorithm.
\end{abstract}

\maketitle

\section{Introduction}

Non-local correlations displayed by certain entangled quantum
systems mark a clear departure from the classical framework made 
of well-definite locally interacting quantities~\cite{bell}. 
Besides their importance in foundation of quantum theory, 
non-local correlations have gained interest as information-processing
resources in cryptography~\cite{barrett0,acin0,scarani,acin2,acin3,masanes,hanggi}, 
randomness amplification~\cite{colbeck,gallego}, quantum computation 
and communication complexity~\cite{buhrman}. In view of their
importance, a relevant problem --- hereafter called
\emph{non-locality problem} --- is to find a
criterion for deciding if observed correlations 
are actually non-local. Such a criterion is for example 
provided by \emph{Bell inequalities}~\cite{pitowski}.
However, a result by Pitowski~\cite{pitowski}
seems to suggest that the problem of discriminating between
local and non-local correlations is generally exponentially
complex. Namely, Pitowski proved 
that deciding the membership to the correlation polytope,
a generalization of the local polytope, is NP-complete and,
therefore, intractable, unless NP is 
equal to P. This result also implies that the opposite problem, 
deciding whether given correlations are outside the polytope,
is not even in NP, unless NP=co-NP, which is 
believed to be false.

In this paper, we briefly revise Pitowski's results and 
their relationship with the problem of discriminating
between local and non-local correlations. We argue that
these results and the NP$\ne$P hypothesis do not have any
obvious implication on the tractability of the non-locality
problem. Falling outside the task of determining
the complexity class of the problem, the purpose of
this work is to present an algorithm which \emph{de facto} 
exhibits polynomial running time in all the performed 
numerical tests. More precisely, the algorithm computes
the distance from the local polytope. First, we prove that
the time cost of computing the distance with arbitrary fixed
error grows polynomially in the size
of the problem input (number of measurements and outcomes),
granted the access to a certain oracle. Namely, given a fixed 
error, we derive two upper bounds on the running time. The first
bound is linear in the number of measurements.
The second bound scales as the number of measurements to the sixth 
power. The former is dominant only for a very high number of measurements 
and is never saturated in the performed numerical tests. Thus, the 
problem of computing the distance is reduced to determining an efficient
simulation of the oracle. Then, we introduce a simple algorithm 
that simulates the oracle. The algorithm is probabilistic and
provides the right answer in a subset of randomized inputs.
Thus, to have a correct answer with sufficiently high probability,
the simulation of the oracle has to be performed with a suitably
high number of initial random inputs. This number is determined
by the desired probability of success and the ratio between the
the cardinality of the subset of inputs providing the 
right answer and the number
of the overall random inputs. As we will not derive this ratio,
the number of random initial trials is pragmatically
chosen in the numerical tests such that the
simulation of the oracle contributes to the overall running time 
with a multiplicative factor and, thus, does not affect
the sixth-power law of the oracle-assisted algorithm. 
In all the performed numerical tests, the overall algorithm 
always computes the distance within the desired accuracy.
The scaling of the running time observed in the tests
is compatible with the sixth-power law derived theoretically.

The paper is organized as follows. In Sec.~\ref{sec_nonsign_box},
we introduce our general scenario. For the
sake of simplicity, we will discuss only the two-party
case, but the results can be extended to the general
case of many parties. After introducing the local polytope
in Sec.~\ref{loca_poly_sec}, we revise the results by Pitowski
in Sec.~\ref{pito_sec}. In Sec.~\ref{sec_distance}, we formulate
the non-locality problem as a minimization problem, namely,
the problem of computing the distance from the local polytope.
In Sec.~\ref{algo_section}, the algorithm is introduced. 
The convergence and the computational 
cost are then discussed in Sec.~\ref{convergence_sec}.
After introducing the algorithm for solving the oracle, we
finally discuss the numerical results in Sec.~\ref{sec_numerical}.

\section{Nonsignaling box}
\label{sec_nonsign_box}
In a Bell scenario, two quantum systems are prepared in an entangled state
and delivered to two spatially separate parties, say Alice and Bob. 
These parties perform each a measurement on their system and
get an outcome. In general, Alice and Bob are allowed to choose among
their respective sets of possible measurements. We assume that the sets 
are finite, but arbitrarily large. Let us denote the measurements 
performed by Alice and Bob by the indices $a\in\{1,\dots,A\}$ and 
$b\in\{1,\dots,B\}$, respectively. 
After the measurements, Alice gets an outcome $r\in {\cal R}$ and Bob an 
outcome $s\in{\cal S}$, where $\cal R$ and $\cal S$ are two sets with
cardinality $R$ and $S$, respectively.
The overall scenario is described by the joint conditional probability 
$P(r,s|a,b)$ of getting $(r,s)$ given $(a,b)$.
Since the parties are spatially separate, causality and 
relativity imply that this distribution satisfies the nonsignaling conditions
\begin{equation}
\begin{array}{c}
\label{ns-conds}
P(r|a,b)=P(r|a,\bar b)\;\; \forall r,a,b,\bar b, 
\vspace{1mm} \\
P(s|a,b)=P(s|\bar a,b)\;\; \forall s,b,a,\bar a,
\end{array}
\end{equation}
where 
$P(r|a,b)\equiv \sum_{s} P(r,s|a,b)$ and $P(s|a,b)\equiv \sum_{r} P(r,s|a,b)$
are the marginal conditional probabilities of $r$ and $s$, respectively.
In the following discussion, we consider a more general scenario than
quantum correlations, and we just assume that $P(r,s|a,b)$ satisfies the 
nonsignaling conditions. The abstract machine producing the correlated 
variables $r$ and $s$ from the inputs $a$ and $b$ will be called 
{\it nonsignaling box} (briefly, NS-box). 

\section{Local polytope}
\label{loca_poly_sec}

The correlations between the outcomes $r$ and $s$ associated with the measurements 
$a$ and $b$ are local if and only if the conditional probability $P(r,s|a,b)$ can
be written in the form
\begin{equation}
\label{local_cond}
P(r,s|a,b)=\sum_x P^A(r|a,x)P^B(s|b,x) P^S(x),
\end{equation}
where $P^A$, $P^B$, and $P^S$ are suitable probability distributions. 
In such a case, we say that the nonsignaling box $P(r,s|a,b)$ is local. It is 
always possible to write the conditional probabilities $P^A$ and $P^B$ as convex 
combination of local deterministic processes, that is,
\begin{equation}
\begin{array}{c}
P^A(r|a,x)=\sum_{\bf r} P^A_{\text{det}}(r|{\bf r},a) \rho^A({\bf r}|x), 
\vspace{1mm} \\
P^B(s|b,x)=\sum_{\bf s} P^B_{\text{det}}(s|{\bf s},b) \rho^B({\bf s}|x),
\end{array}
\end{equation}
where ${\bf r}\equiv(r_1,\dots,r_A)$, ${\bf s}\equiv(s_1,\dots,s_B)$, 
$P^A_{\text{det}}(r|{\bf r},a)=\delta_{r_a,r}$ and
$P^B_{\text{det}}(s|{\bf s},b)=\delta_{s_b,s}$.
Using this decomposition, Eq.~(\ref{local_cond}) takes the form of a
convex combination of local deterministic distributions. That is,
\begin{eqnarray}
\nonumber
P(r,s|a,b) &=& \sum_{{\bf r},{\bf s}} 
P^A_{\text{det}}(r|{\bf r},a) P^B_{\text{det}}(s|{\bf s},b) P^{AB}({\bf r},{\bf s}) \\
\nonumber
&=& \sum_{{\bf r},{\bf s}} \delta_{r,r_a} \delta_{s,s_b} P^{AB}({\bf r},{\bf s}) \\
\label{local_distr}
&=& \sum_{{\bf r},r_a=r} \sum_{{\bf s},s_b=s} P^{AB}({\bf r},{\bf s}).
\end{eqnarray}
where 
$P^{AB}({\bf r},{\bf s})\equiv\sum_x \rho^A({\bf r}|x) \rho^B({\bf s}|x) P^S(x)$ and
$\delta_{i,j}$ is the Kronecker delta. Eq.~(\ref{local_distr}) is known
as Fine's theorem~\cite{fine}.
Thus, a local distribution can always be written as convex combination of 
local deterministic distributions. Clearly, the converse is also true
and a convex combination of local deterministic distributions is local.
Therefore,
the set of local distributions is a polytope, called \emph{local polytope}.
As the deterministic probability distributions 
$P^A_{\text{det}}(r|{\bf r},a)P^B_{\text{det}}(s|{\bf s},b)$ are not convex 
combination of other distributions, they all define the vertices of the local 
polytope. Thus, there are $R^A S^B$ vertices, each one being specified by the 
sequences 
${\bf r}$  and $\bf s$. Let us denote the map from $({\bf r},{\bf s})$
to the associated vertex by $\vec V$. That is, $\vec V$ maps the sequences 
to a deterministic local distribution,
\begin{equation}
\vec V({\bf r},{\bf s})\equiv \left[(r,s,a,b)\rightarrow \delta_{r,r_a}
\delta_{s,s_b}\right].
\end{equation}
Since the elements of the
local polytope are normalized distributions and satisfy the nonsignaling
conditions~(\ref{ns-conds}), the $R S A B$ parameters defining $P(r,s|a,b)$
are not independent and the polytope lives in a 
lower-dimensional subspace. The dimension of this subspace and, more 
generally, of the subspace of NS-boxes is equal to~\cite{collins}
\begin{equation}
d_{NS}\equiv A B(R-1)(S-1)+A(R-1)+B(S-1).
\end{equation}

By the Minkowski-Weyl theorem,
the local polytope can be represented as the intersection of finitely many half-spaces.
A half-space is defined by an inequality
\begin{equation}
\label{bell_ineq}
\sum_{r,s,a,b}P(r,s|a,b)B(r,s;a,b)\le L.
\end{equation}
In the case of the local polytope, these inequalities are called \emph{Bell inequalities}.
Given the coefficients $B(r,s;a,b)$, we can choose $L$ such that the inequality
is as restrictive as possible. This is attained by imposing that at least
one vertex of the local polytope is at the boundary of the half-space, that is,
by taking
\begin{equation}
L=\max_{{\bf r},{\bf s}}\sum_{a,b}B(r_a,s_b;a,b).
\end{equation}
The oracle, which is central in this work and introduced later in 
Sec.~\ref{sec_distance}, returns the value $L$ from the coefficients
$B(r,s;a,b)$.

A minimal representation of a polytope is given by the set of facets of the
polytope. A half-space $\sum_{r,s,a,b}P(r,s|a,b)B(r,s;a,b)\le L$
specifies a facet if the associated hyperplane 
$\sum_{r,s,a,b}P(r,s|a,b)B(r,s;a,b)=L$ intersects the boundary of
the polytope in a set with dimension equal to the dimension of the
polytope minus one. A distribution $P(r,s|a,b)$ is local if and only if
every facet inequality is not violated. To check the violation
of every inequality is generally believed to be intractable because of
a result by Pitowski~\cite{pitowski}, but to test 
the membership of a distribution to the local polytope can be done in 
polynomial time, once the vertices of which the distribution is a
convex combination are known. Thus, deciding the membership to the
local polytope is an NP problem. In the following section, we will 
discuss what is actually known about the complexity of the non-locality 
problem.

\section{Correlation polytope and computational complexity}
\label{pito_sec}
A local polytope is a particular case of a correlation polytope, the latter
have been extensively studied by Pitowski in the context of quantum
theory~\cite{pitowski}. In particular, Pitowski proved that
deciding the membership to the correlation polytope is NP-complete.
Here, we shortly revise this result and we argue that it does not
directly imply that deciding the membership to the local polytope
is also NP-complete. For the sake of consistence with the previous
definition of local polytope, we will use slightly different
but equivalent notations with respect to the ones in 
Ref.~\cite{pitowski}.

A correlation polytope is introduced in the following scenario.
There is a set of $M$ possible measurements or queries, labelled
by an index $m=1,\dots,M$. Each query returns a binary outcome
in $\{0,1\}$. Let $K_M$ be the set of pairs $(a,b)$ with 
$a,b\in \{1,\dots,M\}$ and $S$ a subset of $K_M$. We denote by
${\cal R}(M,S)$ the space of real functions 
$f:\{0,1\}^2 \times S\rightarrow \mathbb{R}$.
A sequence ${\bf w}\in \{0,1\}^M$ yields a function $v_{\bf w}\in
{\cal R}(M,S)$ as follows,
\begin{equation}
v_{\bf w}(r,s|a,b)\equiv \delta_{r,w_a}\delta_{s,w_b}, \;\; 
(a,b)\in S,\;\; r,s\in \{0,1\}.
\end{equation}
A function $v_{\bf w}$ can be interpreted as a deterministic
probability distribution that assigns probability $1$ to the 
outcomes $(r=w_a,s=w_b)$ given a pair of queries $(a,b)\in S$.
In the case that
\begin{equation}
S=\{(a,b);a\in {\cal A},b\in {\cal B}\}\equiv S_{\text{loc}},
\end{equation}
where $\cal A$ and $\cal B$ are
disjoint subsets of $\{1,\dots,M\}$, the functions
$v_{\bf w}$ correspond to vertices of a local polytope, as
defined in Sec.~\ref{loca_poly_sec}.
\begin{definition}
A correlation polytope $c(M,S)\subset {\cal R}(M,S)$ is the convex 
hull of the functions $v_{\bf w}$.
\end{definition}
A correlation polytope is a local polytope if $S=S_{\text{loc}}$.
Pitowski proved the following~\cite{pitowski}.
\begin{theorem}
Deciding the membership to a correlation polytope is
NP-complete.
\end{theorem}
For this task, Pitowski considers particular instances of
the problem with $S=S_{k,M}$, the set $S_{k,M}$ containing all
pairs $(a,b)$ with $1 \le a<b\le M$ except the pairs $\{1,M\}$,
$\{2,M\},\dots,\{k,M\}$. These instances are then reduced
to the one-in-three 3-satisfiability problem, which is NP-complete.
It is however clear that the set $S_{k,M}$ does not take the
form of $S_{\text{loc}}$. The latter set is made of pairs of elements
in disjoint sets, whereas the set $S_{k,M}$ contains almost
every pair except $\{1,M\},\dots,\{k,M\}$. 

In conclusion, we argue that Pitowski's theorem does not imply that 
deciding the membership to a local polytope is NP-complete. It is possible
that Pitowski's argument could be extended to include also the case 
of local polytopes, but this extension is not obvious. In this 
paper, we will not provide an answer to the question, but we will
present an algorithm that computes efficiently the distance from the 
local polytope in all the physically relevant cases numerically
tested. The question whether the algorithm has always polynomial running 
time is left as an open problem.

\section{Distance from the local polytope}
\label{sec_distance}
The non-locality problem can be reduced to a convex optimization problem,
such as the computation of the nonlocal capacity, introduced in 
Ref.~\cite{montina}, and the distance from the local polytope,
which can be reduced to a linear program if the $L^1$ norm is
employed~\cite{bernhard}.
Here, we define the distance of a distribution 
$P(r,s|a,b)$ from the local polytope as the Euclidean distance between 
$P(r,s|a,b)$ and the closest local distribution. As said in Sec.~\ref{loca_poly_sec} 
[see Eq.~(\ref{local_distr})] and stated by Fine's 
theorem~\cite{fine}, a conditional distribution $\rho(r,s|a,b)$ is local if and 
only if there is a non-negative function $\chi({\bf r},{\bf s})$ such that
\begin{equation}
\label{local_cond_prob}
\rho(r,s|a,b)=\sum_{{\bf r},r_a=r}\sum_{{\bf s},s_b=s}\chi({\bf r},{\bf s}).
\end{equation}
That is, a conditional distribution $\rho(r,s|a,b)$ is local if it is 
the marginal of a multivariate probability distribution $\chi$ of the
outcomes of all the possible measurements, provided the $\chi$ does
not depend on the measurements $a$ and $b$. 

The distributions $P(r,s|a,b)$ and $\rho(r,s|a,b)$ can be represented
as vectors in a $R S A B$-dimensional space. Let us denote them by
$\vec P$ and $\vec\rho$, respectively. 
Given a positive-definite matrix $\hat M$ defining the metrics
in the vector space, the computation of the distance from the local 
polytope is equivalent to the minimization of a functional of the form
\begin{equation}
F[\chi]=\frac{1}{2}\left(\vec P-\vec \rho\right)^T\hat M
\left(\vec P-\vec \rho\right)
\end{equation}
with respect to $\chi$, under the constraints that $\chi$ is non-negative
and normalized. Namely, the distance is the square root of the minimum of 
$2 F$. Hereafter, we choose the metrics so that the functional
takes the form
\begin{equation}
F[\chi]\equiv \frac{1}{2} \sum_{r,s,a,b}\left[P(r,s|a,b)-\rho(r,s|a,b)\right]^2
W(a,b),
\end{equation}
where $W(a,b)$ is some probability distribution. The normalization
$\sum_{a,b}W(a,b)=1$ guarantees that the distance does not diverge
in the limit of infinite measurements performed on a given entangled 
state. In particular, we will consider the case with 
\begin{equation}
W(a,b)\equiv \frac{1}{A B}.
\end{equation}
Another choice would be to take the distribution $W(a,b)$ maximizing
the functional, so that the computation of the distance would be a minimax 
problem. This case has some interesting advantages, but is more sophisticated  
and will not be considered here. Since we are interested in a quantity that is 
equal to zero if and only if $P(r,s|a,b)$ is local, we can simplify the 
problem of computing the distance by dropping the normalization constraint
on $\chi$. Indeed, if the distance is equal to zero, $\rho$ and,
thus, $\chi$ are necessarily normalized. Conversely, if the distance
is different from zero for every normalized local distribution, it
is so also for every unnormalized local distribution. Thus, the
discrimination between local and non-local correlation is equivalent to
the following minimization problem.
\begin{problem}
\label{main_problem}
\begin{equation}
\nonumber
\begin{array}{c}
\min_\chi F[\chi] \\
\text{subject to the constraints} \\
\chi({\bf r},{\bf s})\ge0.
\end{array}
\end{equation}
\end{problem}
Let us denote the solution of this problem and the corresponding
optimal value by $\chi^{min}$ and $F^{min}$, respectively. The
associated (unnormalized) local distribution is denoted by
$\rho^{min}(r,s|a,b)$.
The square root of $2 F^{min}$ is the minimal distance of $P(r,s|a,b)$
from the cone defined as the union of all the lines connecting 
the zero distribution $\rho(r,s|a,b)=0$ and an arbitrary point
of the local polytope. Let us call this set \emph{local cone}.
Hereafter, we will consider the problem of computing the distance 
from the local cone, but the results can be easily extended to the 
case of the local polytope, so that we will use ``local cone''
and ``local polytope'' as synonyms in the following discussion.
Note that there are generally infinite minimizers $\chi^{min}$, since
$\chi$ lives in a $R^A S^B$-dimensional space, whereas the
functional $F$ depends on $\chi$ through $\rho(r,s|a,b)$, which
lives in a $(d_{NS}+1)$-dimensional space.
In other words, since the local polytope has $R^A S^B$
vertices, but the dimension of the polytope is $d_{NS}$, a
(normalized) distribution $\rho$ has generally infinite representations
as convex combination of the vertices, unless $\rho$ is on a face whose 
dimension plus $1$ is equal to the number of vertices defining
the face.

At first glance, the computational complexity of this problem seems 
intrinsically exponential, as the number of real variables
defining $\chi$ is equal to $R^A S^B$. However, the dimension of
the local polytope is $d_{NS}$ and grows polynomially in the
number of measurements and outcomes. Thus, by Carath\'eodory's
theorem, a (normalized) local distribution can always be represented 
as the convex combination of a number of vertices smaller
than $d_{NS}+2$. This implies that there is a minimizer
$\chi^{min}$ of $F$ whose support contains a number of
elements not greater than $d_{NS}+1$. Therefore, the minimizer
can be represented by a number of variables growing polynomially 
in the input size. The main problem  is to find a small set 
of vertices that are suitable for representing the closest local 
distribution $\rho^{min}(r,s|a,b)$. In the following,
we will show that the computation of the distance from
the local cone with arbitrary fixed accuracy has polynomial 
complexity, granted the access to the following oracle.
\newline
{\bf Oracle}: Given a function $g(r,s;a,b)$, the 
oracle returns the sequences $\bf r$ and $\bf s$ maximizing the 
function
\begin{equation}
G({\bf r},{\bf s})\equiv \sum_{a,b}g(r_a,s_b;a,b) W(a,b)
\end{equation}
and the corresponding maximal value.\newline
Thus, Problem~\ref{main_problem} is reduced to determining
an efficient simulation of the oracle. Let us
consider the case of binary outcomes with 
$r$ and $s$ taking values $\pm1$ ($R=S=2$). 
The function $G({\bf r},{\bf s})$ takes the form
\begin{equation}
\label{spin_glass}
G({\bf r},{\bf s})=\sum_{a,b}J_{ab}r_a s_b+\sum_a A_a r_a+\sum_b B_b s_b+G_0,
\end{equation}
whose minimization falls into the class of spin-glass problems,
which are notoriously computationally
hard to handle. This would suggest that the oracle is
generally an intractable problem. Nonetheless, the oracle
has a particular structure that can make the problem
easier to be solved. This will be discussed later in 
Secs.~\ref{sec_oracle} and \ref{sec_numerical}. There, we 
will show that the 
oracle can be simulated efficiently in many relevant cases by
using a simple block-maximization strategy. 
Assuming for the moment that we have access to the oracle, let 
us introduce the algorithm solving Problem~\ref{main_problem}.

\section{Computing the distance}
\label{algo_section}

The distance from the local polytope can be computed efficiently
once we have a set $\Omega$ of vertices that is small enough and suitable 
for representing the closest distribution $\rho^{min}(r,s|a,b)$.
The algorithm introduced in this paper solves Problem~\ref{main_problem} 
by generating iteratively a sequence of sets $\Omega$. At each step, 
the minimal distance is first computed over the convex hull of the given 
vertices. Then, the oracle is consulted. As well as the set does not contain 
the right vertices, the oracle returns a strictly positive maximal value 
and a vertex, which is added to the set $\Omega$ (after possibly removing
vertices with zero weight). The optimization
Problem~\ref{main_problem} is solved once the oracle returns zero,
which guarantees that
all the optimality conditions of the problem are satisfied. 
Before discussing the algorithm, let us derive these conditions.

\subsection{Necessary and sufficient conditions for optimality}
Problem~\ref{main_problem} is a convex optimization problem
whose constraints satisfy Slater's condition, requiring
the existence of an interior point of the feasible region.
This is the case, as a positive $\chi$ strictly satisfies all 
the inequality constraints. Thus, the four Karush--Kuhn--Tucker 
conditions are necessary and sufficient conditions for 
optimality. The first
requires that the gradient of the Lagrangian is equal to
zero (stationarity condition). The Lagrangian of 
Problem~\ref{main_problem} is
\begin{equation}
\label{lagrangian}
{\cal L}=F[\chi]-\sum_{{\bf r},{\bf s}}\lambda({\bf r},{\bf s})
\chi({\bf r},{\bf s}),
\end{equation}
where $\lambda({\bf r},{\bf s})$ are the Lagrange multipliers
associated with the inequality constraints. The second condition
is the feasibility of the constraints. The third condition,
called \emph{dual feasibility}, is the non-negativity of
$\lambda$, that is,
\begin{equation}
\lambda({\bf r},{\bf s})\ge 0.
\end{equation}
Finally, the last condition, the \emph{complementary slackness},
states that
\begin{equation}
\lambda({\bf r},{\bf s})\chi({\bf r},{\bf s})=0.
\end{equation}
The stationarity condition on the gradient of the Lagrangian
gives the equality
\begin{equation}
\label{slackness0}
\sum_{a,b}W(a,b)\left[P(r_a,s_b|a,b)-\rho(r_a,s_b|a,b)\right]+
\lambda({\bf r},{\bf s})=0.
\end{equation}
Eliminating $\lambda$,
this equality and the dual feasibility yield the inequality
\begin{equation}
\label{cond_1}
\sum_{a,b}W(a,b)\left[P(r_a,s_b|a,b)-\rho(r_a,s_b|a,b)\right]\le 0.
\end{equation}
From Eq.~(\ref{slackness0}), we have that
the complementary slackness is equivalent to the following
condition,
\begin{equation}
\label{cond_2}
\begin{array}{c}
\chi({\bf r},{\bf s})\ne0\Rightarrow 
\vspace{1mm} \\
\sum_{a,b}W(a,b)\left[P(r_a,s_b|a,b)-\rho(r_a,s_b|a,b)\right]=0,
\end{array}
\end{equation}
that is, the left-hand side of the last inequality 
is equal to zero if $({\bf r},{\bf s})$ is in
the support of $\chi$. The slackness condition~(\ref{cond_2}),
the primal constraint 
and Ineq.~(\ref{cond_1}) provide necessary and sufficient
conditions for optimality. Let us introduce the function
\begin{equation}
g(r,s;a,b)\equiv P(r,s|a,b)-\rho(r,s|a,b),
\end{equation}
which is the opposite of the gradient of $F$ with respect to 
$\rho$ up to the factor $W(a,b)$. Summarizing, the conditions 
are
\begin{eqnarray}
\label{first_cond}
&\sum_{a,b} W(a,b) g(r_a,s_b;a,b)\le 0,& \\
\label{second_cond}
&\chi({\bf r},{\bf s})\ne0\Rightarrow 
\sum_{a,b} W(a,b) g(r_a,s_b;a,b)=0,& \\
\label{third_cond}
& \chi({\bf r},{\bf s})\ge0. &
\end{eqnarray}
The second condition can be rewritten in the more concise
form
\begin{equation}
\label{slackness1}
\sum_{r,s,a,b} \rho(r,s|a,b) g(r,s|a,b) W(a,b)=0.
\end{equation}
Indeed, using Ineqs.~(\ref{first_cond},\ref{third_cond}),
it is easy to show that condition~(\ref{second_cond}) is 
satisfied if and only if
$$
\sum_{{\bf r},{\bf s}}\chi({\bf r},{\bf s})
\sum_{a,b} W(a,b) g(r_a,s_b;a,b)=0,
$$
which gives equality~(\ref{slackness1})
by definition of $\rho$ [Eq.~(\ref{local_cond_prob})].

Condition~(\ref{first_cond}) can be checked by
consulting the oracle with $g(r,s;a,b)$ as query. If
the oracle returns a non-positive maximal value, then
the condition is satified. Actually, at the optimal
point, the returned value turns out to be equal to
zero, as implied by the other optimality conditions.

Similar optimality conditions hold if we force $\chi$ 
to be equal to zero outside some set $\Omega$. Let us
introduce the following minimization problem.
\begin{problem}
\label{secondary_problem}
\begin{equation}
\nonumber
\begin{array}{c}
\min_\chi F[\chi] \\
\text{subject to the constraints} \\
\chi({\bf r},{\bf s})\ge0, \\
\chi({\bf r},{\bf s})=0 \;\;\; \forall ({\bf r},{\bf s})\notin \Omega 
\end{array}
\end{equation}
\end{problem}
The optimal value of this problem gives an upper bound on the optimal
value of Problem~\ref{main_problem}. The two problems are equivalent
if the support of a minimizer $\chi^{min}$ of 
Problem~\ref{main_problem} is in $\Omega$. The necessary and
sufficient conditions for optimality of Problem~\ref{secondary_problem}
are the same as of Problem~\ref{main_problem}, with the only difference
that condition~(\ref{first_cond}) has to hold only in the set $\Omega$.
That is, the condition is replaced by the weaker condition
\begin{equation}
\label{first_cond_prob2}
({\bf r},{\bf s})\in\Omega \Rightarrow
\sum_{a,b} W(a,b) g(r_a,s_b;a,b)\le 0.
\end{equation}
Thus, an optimizer of Problem~\ref{secondary_problem} is solution
of Problem~\ref{main_problem} if the value returned by the oracle
with query $g=P-\rho$ is equal to zero.

Hereafter, the minimizer and the minimal value of Problem~\ref{secondary_problem}
will be denoted by $\chi^{min}_\Omega$ and $F^{min}_\Omega$, respectively.
The associated optimal local distribution $\rho(r,s|a,b)$ defined by 
Eq.~(\ref{local_cond_prob}) will be denoted by $\rho^{min}_\Omega(r,s|a,b)$.

\subsection{Overview of the algorithm}
\label{algo_overview}
Problem~\ref{main_problem} can be solved iteratively by finding the solution 
of Problem~\ref{secondary_problem} over a sequence of sets $\Omega$. The 
sets are built according to the answer of the oracle, which is consulted at 
each step of the iteration. The procedure stops when a desired accuracy is
reached or $\Omega$ contains the 
support of a minimizer $\chi^{min}$ and the solution of Problem~\ref{secondary_problem}
is also solution of Problem~\ref{main_problem}. Let us outline the algorithm.
Suppose that we choose the initial $\Omega$ as a set of sequences 
$({\bf r},{\bf s})$ associated to $n_0$ linearly independent vertices
($n_0$ is possibly equal to $1$).
Let us denote this set by $\Omega_0$. We solve Problem~\ref{secondary_problem}
with $\Omega=\Omega_0$
and we get the optimal value $F_0^{min}\equiv F^{min}_{\Omega_0}$ 
with minimizer $\chi_0^{min}\equiv \chi^{min}_{\Omega_0}$.
Let us denote the corresponding (unnormalized) local distribution by 
$\rho_0^{min}\equiv \rho^{min}_{\Omega_0}$.
That is,
\begin{equation}
\rho_0^{min}(r,s|a,b)\equiv\sum_{{\bf r},r_a=r}\sum_{{\bf s},s_b=s}\chi_0^{min}({\bf r},{\bf s}).
\end{equation}
Since the cardinality
of $\Omega_0$ is not greater than $d_{NS}+1$ and the problem is a convex quadratic 
optimization problem, the corresponding computational complexity is polynomial.
Generally, a numerical algorithm provides an optimizer up to some arbitrarily
small but finite error. 
In Sec.~\ref{sec_stop_crit}, we will provide a bound on the accuracy required
for the solution of Problem~\ref{secondary_problem}. For now, let
us assume that Problem~2 is solved exactly. If the support of $\chi^{min}$ is 
in $\Omega_0$, $F_0^{min}$ is equal to the optimal value of 
Problem~\ref{main_problem}, and we have computed the distance from the local 
polytope. We can verify if this is the case by checking the first optimality 
condition~(\ref{first_cond}), as the conditions~(\ref{second_cond},\ref{third_cond})
are trivially satisfied by the optimizer of Problem~\ref{secondary_problem}
for every $({\bf r},{\bf s})$. The check is made by consulting the oracle 
with the function $P(r,s|a,b)-\rho_0^{min}(r,s|a,b)$ as query. If the oracle returns 
a maximal value equal to zero, then we 
have the solution of Problem~\ref{main_problem}. Note that if the optimal 
value of Problem~\ref{secondary_problem} is equal to zero, then also the
optimal value of the main problem is equal to zero and the conditional
distribution $P(r,s|a,b)$ is local. In this case, we have no need
of consulting the oracle.

If the optimal value of Problem~\ref{secondary_problem} is different
from zero and the oracle returns a maximal value strictly positive,
then the minimizer of Problem~\ref{secondary_problem} satisfies all 
the optimality conditions of Problem~\ref{main_problem}, except 
Ineq.~(\ref{first_cond}) for some $({\bf r},{\bf s})\notin\Omega$. 
The next step is to add the pair of sequences $({\bf r},{\bf s})$
returned by the oracle to the set $\Omega$ and solve 
Problem~\ref{secondary_problem} with the new set. Let us denote
the new set and the corresponding optimal value by $\Omega_1$
and $F_1^{min}\equiv F^{min}_{\Omega_1}$, respectively. 
Once we have solved Problem~\ref{secondary_problem} with $\Omega=\Omega_1$,
we consult again the oracle to check if we have obtained the solution of 
Problem~\ref{main_problem}. If we have not, we add the pair of sequences
$({\bf r},{\bf s})$ given by the oracle to the set $\Omega$ and we
solve Problem~\ref{secondary_problem} with the new set, say $\Omega_2$.
We continue
until we get the solution of Problem~\ref{main_problem} or its optimal
value up to some desired accuracy. This procedure generates a sequence of
sets $\Omega_{n=1,2,\dots}$ and values $F^{min}_{n=1,2,\dots}$. The latter
sequence is strictly decreasing, that is, $F_{n+1}^{min}<F_n^{min}$ until $\Omega_n$
contains the support of $\chi^{min}$ and the oracle returns zero as
maximal value. Let us show that. Suppose that $\chi_n^{min}$ 
is the optimizer of Problem~\ref{secondary_problem} with $\Omega=\Omega_n$
and $({\bf r}',{\bf s}')$ is the new element in the set $\Omega_{n+1}$.
Let us denote by $\rho_n^{min}(r,s|a,b)$ the local distribution associated
with $\chi_n^{min}$, that is,
\begin{equation}
\label{rho_n}
\rho_n^{min}(r,s|a,b)\equiv\sum_{{\bf r},r_a=r}\sum_{{\bf s},s_b=s}\chi_n^{min}({\bf r},{\bf s}).
\end{equation}
The optimal value $F_{n+1}^{min}$ of Problem~\ref{secondary_problem} is bounded
from above by the value taken by the function $F[\chi]$ for every
feasible $\chi$, in particular, for 
\begin{equation}
\chi({\bf r},{\bf s};\alpha)=\chi_n^{min}({\bf r},{\bf s})+\alpha \delta_{{\bf r},{\bf r}'}
\delta_{{\bf s},{\bf s}'},
\end{equation}
with $\alpha$ positive. Let us set $\alpha$ equal to the value
minimizing $F$, that is,
\begin{equation}
\label{alpha_n}
\alpha\equiv\alpha_n=
\sum_{a b}W(a,b)[P(r_a',s_b'|a,b)-\rho_n^{min}(r_a',s_b'|a,b)],
\end{equation} 
which is equal to the value returned by the oracle.
It is strictly positive, as the oracle returned a positive value,
provided that $\Omega_n$ does not contain the support of $\chi^{min}$.
Hence, $\chi({\bf r},{\bf s};\alpha_n)$ is a feasible point and, thus, 
the corresponding value taken by $F$,
\begin{equation}
\label{upper_b_F}
\left. F\right|_{\alpha=\alpha_n}=F_n^{min}-\frac{1}{2} \alpha_n^2,
\end{equation}
is an upper bound on $F_{n+1}^{min}$. Hence,
\begin{equation}
\label{sequence_F}
F_{n+1}^{min}\le  F_n^{min}-\frac{1}{2} \alpha_n^2,
\end{equation}
that is, $F_{n+1}^{min}$ is strictly smaller than $F_n^{min}$.

This procedure generates a sequence $F_n^{min}$ that converges to 
the optimal value of Problem~\ref{main_problem}, as shown in 
Sec.~\ref{convergence_sec}. For any given accuracy, the 
computational cost of the procedure is polynomial, provided that 
we have access to the oracle.

To avoid a growth of the cardinality of $\Omega$ beyond $d_{NS}+1$ during 
the iteration and, thus, the introduction of redundant vertices, we have to 
be sure that the sets $\Omega_0, \Omega_1,\dots$ contain points $({\bf r},{\bf s})$
associated to linearly independent vertices $\vec V({\bf r},{\bf s})$
of the local polytope. This is guaranteed
by the following procedure of cleaning up. First, after the computation of
$\chi_n^{min}$ at step $n$, we remove the elements in $\Omega_n$ where
$\chi_n^{min}({\bf r},{\bf s})$ is equal to zero. Then, the
set $\Omega_{n+1}$ is built by adding the point given by the oracle
to the remaining elements. Let us denote by $\cal V$
the set of vertices associated to the elements in the support of $\chi_n^{min}$. 
The cleaning up ensures that the optimizer $\rho_n^{min}$ is in the interior 
of the convex hull of $\cal V$, up to a normalization constant, and
the new vertex returned by the oracle is linearly independent of the ones
in $\cal V$.
Indeed, we have seen that the introduction of such a vertex
allows us to lower the optimal value of Problem~\ref{secondary_problem}.
This would not be possible if the added vertex was linearly
dependent on the vertices in $\cal V$, as the (normalized) optimizer 
$\rho_n^{min}$ of Problem~\ref{secondary_problem} is in the interior of the 
convex hull of $\cal V$. 

This is formalized in Lemma~1.
\begin{lemma}
Let $({\bf r}',{\bf s}')$ be a sequence such that 
\begin{equation}
\label{hypo1}
\sum_{a,b} g(r_a',s_b';a,b) W(a,b)\ne 0.
\end{equation}
If $\Omega$ is a set of sequences such that
\begin{equation}
\label{gradient_zero}
({\bf r},{\bf s})=\Omega \Rightarrow 
\sum_{a,b} g(r_a,s_b;a,b) W(a,b)=0, 
\end{equation}
then the vertex $\vec V({\bf r}',{\bf s}')$ is linearly independent
of the vertices associated to the sequences in $\Omega$.
\end{lemma}
{\it Proof}.
The proof is by contradiction. Suppose that the vector $\vec V({\bf r}',{\bf s}')$
is linearly dependent of the vectors $\vec V({\bf r},{\bf s})$ with 
$({\bf r},{\bf s})\in\Omega$, then there is a real function $t({\bf r},{\bf s})$
such that
\begin{equation}
\label{lin_combi}
\vec V({\bf r}',{\bf s}')=
\sum_{({\bf r},{\bf s})\in\Omega} t({\bf r},{\bf s}) \vec V({\bf r},{\bf s}).
\end{equation}
By definition of $\vec V$, this equation implies that
$\sum_{{\bf r},{\bf s}} t({\bf r},{\bf s})\delta_{r,r_a}\delta_{s,s_b}
=\delta_{r,r_a'}\delta_{s,s_b'}$. 
From this equation and Eq.~(\ref{gradient_zero}), we have
\begin{equation}
\sum_{r,s} \delta_{r,r_a'}\delta_{s,s_b'} \sum_{a,b} g(r,s;a,b) W(a,b)
=0.
\end{equation}
Summing over $r$ and $s$, we get a contradiction with Eq.~(\ref{hypo1}).
$\square$ \newline
This lemma and the optimality 
conditions~(\ref{first_cond},\ref{second_cond})
imply that the sets $\Omega_0, \Omega_1,\dots$ built through the
previously discussed procedure of cleaning up always contain
points associated to independent vertices and, thus, never contain 
more than $d_{NS}+1$ elements. Indeed, the previously defined
set $\cal V$ contains points $({\bf r},{\bf s})$ where the 
minimizer $\chi_n^{min}$ is different from zero, for which the quantity 
$\sum_{a,b}\left[P(r_a,s_b|a,b)-\rho_n^{min}(r_a,s_b|a,b)\right]W(a,b)=0$,
as implied by condition~(\ref{second_cond}).
Furthermore, given the sequence $({\bf r}',{\bf s}')$ returned by
the oracle, condition~(\ref{first_cond}) implies that
$\sum_{a,b}\left[P(r_a',s_b'|a,b)-\rho_n^{min}(r_a',s_b'|a,b)\right]W(a,b)>0$
until the set $\Omega_n$ contains the support of $\chi^{min}$ and 
the iteration generating the sequence of sets $\Omega$ is terminated.

The procedure of cleaning up is not strictly 
necessary for having a polynomial running time, but it can 
speed up the algorithm.

\subsection{The algorithm}

In short, the algorithm for computing the distance from the local polytope 
with given accuracy is as follows.
\begin{algorithm}
\label{main_algo}
Input: $P(r,s|a,b)$
\begin{enumerate}
\item Set $({\bf r}',{\bf s}')$ equal to the sequences
given by the oracle with $P(r,s|a,b)$ as query.
\item Set $\Omega=\{ ({\bf r}',{\bf s}') \}$.
\item \label{step_prob2}
Compute the optimizers $\chi({\bf r},{\bf s})$ and 
$\rho(r,s|a,b)$ of Problem~2. The associated $F$
provides an upper bound of the optimal value $F^{min}$.
\item \label{step_oracle}
Consult the oracle with $g(r,s;a,b)=P(r,s|a,b)-\rho(r,s|a,b)$ as query.
Set $({\bf r}',{\bf s}')$ and $\alpha$ equal to the sequences
returned by the oracle and the associated maximal value, 
respectively. That is,
$$
({\bf r}',{\bf s}')=\underset{({\bf r},{\bf s})}{\text{argmax}}
\sum_{a,b} g(r_a,s_b;a,b) W(a,b),
\vspace{-5mm}
$$
$$
\alpha=
\sum_{a,b} g(r_a',s_b'|a,b) W(a,b),
$$
\item 
\label{step_lower_bound}
Compute a lower bound on the $F^{min}$ from $\rho$ and $\alpha$
(see following discussion and Sec.~\ref{sec_dual}).
The difference between the upper and lower bounds provides
an upper bound on the reached accuracy.
\item
\label{step_stop}
If a given accuracy is reached, stop.
\item 
\label{removal_step}
Remove from $\Omega$ the points where $\chi$ is zero
and add $({\bf r}',{\bf s}')$.
\item Go back to Step~\ref{step_prob2}.
\end{enumerate}
\end{algorithm}

The algorithm stops at Step~\ref{step_stop} when a desired
accuracy is reached. To estimate the accuracy, we need to
compute a lower bound on the optimal value $F^{min}$.
To guarantee that the algorithm eventually stops,
the lower bound has to converge to the optimal value
as the algorithm approaches the solution of Problem~1.
We also need a stopping criterion for the numerical routine
solving the optimization problem at Step~\ref{step_prob2}.
Let us first discuss the stopping criterion for
Algorithm~\ref{main_algo}.

\subsection{Stopping criterion for Algorithm~\ref{main_algo}}

The lower bound on $F^{min}$, denoted by $F^{(-)}$, is computed 
by using the dual form of Problem~1. As shown in 
Sec.~\ref{sec_dual}, any local distribution $\rho$ induces
the lower bound
\begin{equation}
\label{lower_bound_0}
F^{(-)}=\frac{1}{2}\sum_{r s a b}\left\{P^2(r,s|a,b)-
\right[\rho(r,s|a,b)+\alpha\left]^2\right\}W(a,b),
\end{equation}
where $\alpha$ is the maximal value returned by the oracle
with $g(r,s;a,b)=P(r,s|a,b)-\rho(r,s|a,b)$ as query. 
An upper bound on $F^{min}$ is obviously
\begin{equation}
F^{(+)}=F[\chi].
\end{equation}
In the limit of $\rho$ equal to the
local distribution minimizing $F$, the lower bound is 
equal to the optimal value $F^{min}$. This can be shown
by using the optimality conditions.
Indeed, conditions~(\ref{first_cond},\ref{slackness1}) 
imply the limits
\begin{equation}
\label{limit1}
\lim_{\chi\rightarrow\chi^{min}}
\alpha=0 , 
\vspace{-3mm}
\end{equation}
\begin{equation}
\label{limit2}
\lim_{\chi\rightarrow\chi^{min}} \sum_{r,s,a,b}\rho(r,s|a,b) g(r,s;a,b) W(a,b)=0,
\end{equation}
which imply
$F^{(-)}\rightarrow F^{min}$ as $\chi$ approaches the
minimizer. This is made more evident by computing the difference
between the upper bound and the lower bound. 
Indeed,
given the local distribution $\rho(r,s|a,b)$ computed at 
Step~\ref{step_prob2} and the corresponding $\alpha$ returned by 
the oracle at Step~\ref{step_oracle}, the difference is
\begin{equation}
\begin{array}{c}
F^{(+)}-F^{(-)}\equiv \Delta F=
\frac{R S}{2}\alpha^2+
\vspace{1mm}  \\
\sum_{r s a b}\rho(r,s|a,b)\left[\alpha-g(r,s;a,b)\right]W(a,b),
\end{array}
\end{equation}
which evidently goes to zero as $\chi$ goes to $\chi^{min}$.
Thus, the upper bound $\Delta F$ on the accuracy computed at 
Step~\ref{step_lower_bound} goes to zero as $\rho(r,s|a,b)$
approaches the solution. This guarantees that the algorithm
stops sooner or later at Step~\ref{step_stop}, provided that
$\chi$ converges to the solution. 
If Problem~\ref{secondary_problem} is solved exactly at
Step~\ref{step_prob2}, then the distribution $\rho(r,s|a,b)$
satisfies condition~(\ref{slackness1}), and the upper bound
on the reached accuracy takes the form
\begin{equation}
F^{(+)}-F^{(-)}=
\frac{R S}{2}\alpha^2+
\alpha \sum_{r s a b}\rho(r,s|a,b) W(a,b).
\end{equation}
Even if Condition~(\ref{slackness1}) is not satisfied, we
can suitably normalize $\chi({\bf r},{\bf s})$ so that
the condition is satisfied. In the following, we assume
that this condition is satisfied.

\subsection{Stopping criterion for Problem~2 (Optimization
at Step~\ref{step_prob2} of Algorithm~\ref{main_algo})}
\label{sec_stop_crit}
In Algorithm~\ref{main_algo}, Step~\ref{step_prob2} is completed
when the solution of Problem~\ref{secondary_problem} with a
given set $\Omega$ is found. Optimization algorithms iteratively
find a solution $\rho^{min}_\Omega(r,s|a,b)$ up to some accuracy. 
We can stop when the error is of the order of the machine precision. 
Here, we will discuss a more effective stopping criterion. 
This criterion should preserve the two main features previously 
described:
\begin{enumerate}
\item
The sequence $F_0^{min}, F_1^{min}, \dots$ of the
exact optimal values of Problem~\ref{secondary_problem}
with $\Omega=\Omega_0,\Omega_1,\dots$ is monotonically decreasing;
\item
The sets $\Omega_0, \Omega_1,\dots$ contain points associated with 
linearly independent vertices of the local polytope, implying that
the cardinality of $\Omega_n$ is never greater than $d_{NS}+1$.
\end{enumerate}
To guarantee that the first feature is preserved, it is sufficient
to compute a lower bound on the optimal value $F^{min}_\Omega$ 
of Problem~\ref{secondary_problem} so that the bound approaches
$F^{min}_\Omega$ as $\chi$ approaches the optimizer $\chi^{min}_\Omega$.
If the lower bound with set $\Omega=\Omega_n$ is greater than 
the upper bound $F_n-\alpha_n^2/2$ on $F^{min}_{n+1}$
[see Eq.~(\ref{upper_b_F})],
then $F^{min}_{n+1}<F^{min}_{n}$. Denoting by $F_{\Omega}^{(-)}$ 
the lower bound on the optimal value $F_\Omega^{min}$,
the monotonicity of the sequence $F_0^{min}, F_1^{min}, \dots$ 
is implied by the inequality
\begin{equation}
\label{ineq_low_bound}
F_n-\frac{1}{2} \alpha_n^2 \le F_{\Omega_n}^{(-)}.
\end{equation}
As shown later by using dual theory, a lower bound on 
$F_\Omega^{min}$ is
\begin{equation}
\label{lower_bound_Omega}
F_\Omega^{(-)}=\frac{1}{2}\sum_{r s a b}\left\{P^2(r,s|a,b)-
\right[\rho(r,s|a,b)+\beta\left]^2\right\}W(a,b),
\end{equation}
where 
\begin{equation}
\beta\equiv \max_{({\bf r},{\bf s})\in\Omega}
\sum_{a b}W(a,b)[P(r_a,s_b|a,b)-\rho(r_a,s_b|a,b)],
\end{equation}
and $\rho(r,s|a,b)$ is an unnormalized local distribution
associated to a function $\chi({\bf r},{\bf s})$ with
support in $\Omega$.
This bound becomes equal to $F_\Omega^{min}$ in the limit
of $\rho$ equal to the minimizer of Problem~2.
Ineq.~(\ref{ineq_low_bound}) gives the condition
\begin{equation}
\label{stop_criterion}
\begin{array}{c}
\alpha^2>
R S \beta^2 +
\vspace{1mm} \\
2\sum_{r s a b}\left[\beta-g(r,s;a,b)\right]\rho(r,s|a,b)
W(a,b),
\end{array}
\end{equation}
where $g(r,s;a,b)=P(r,s|a,b)-\rho(r,s|a,b)$ and $\rho(r,s|a,b)$
is the local distribution computed at Step~3.
If this condition is satisfied by the numerical solution
found at Step~3, then the series $F_0^{min}, F_1^{min},\dots$
is monotonically decreasing. As we will see, to prove
that the series converges to the minimizer of 
Problem~\ref{main_problem}, we need the stronger condition
\begin{equation}
\label{stop_criterion_strong}
\begin{array}{c}
\gamma \alpha^2\ge
R S \beta^2 +
\vspace{1mm} \\
2\sum_{r s a b}\left[\beta-g(r,s;a,b)\right]\rho(r,s|a,b)
W(a,b),
\end{array}
\end{equation}
where $\gamma$ is any fixed real number in the interval
$(0,1)$. A possible choice is $\gamma=1/2$. If this inequality
is satisfied at each iteration of Algorithm~\ref{main_algo},
the sequence $F_0^{min}, F_1^{min},\dots$ satisfies the
inequality
\begin{equation}
\label{num_monotone}
F_{n+1}^{min}\le
F_n^{min}-\frac{1-\gamma}{2}\alpha_n^2,
\end{equation}
which turns out to be equal to Ineq.~(\ref{sequence_F}) in
the limit $\gamma\rightarrow 0$.
The right-hand side of Ineq.~(\ref{stop_criterion_strong})
goes to zero as $\rho$ approaches
the optimizer, as implied by the optimality conditions
of Problem~2. Thus, if the set $\Omega$ does not contain
all the points where $\chi^{min}$ is different from zero,
then the inequality is surely satisfied at some point of the 
iteration solving Problem~2, as $\alpha$ tends to a strictly
positive number. When the inequality is satisfied, the minimization
at Step~3 of Algorithm~\ref{main_algo} is terminated.
If $\Omega$ is the support of $\chi^{min}$, the
inequality will never be satisfied and the minimization
at Step~3 will terminate when the desired accuracy
on $F^{min}$ is reached. 

As previously said, we should also guarantee that the
sets $\Omega_n$ contain only points
associated with linearly independent vertices. This is
granted if the procedure at Step~\ref{removal_step} of
Algorithm~\ref{main_algo} successfully removes the points
where the exact minimizer $\chi_n^{min}$ is equal to zero.
How can we find the support of the minimizer from the
approximate numerical solution computed at Step~3? 
Using dual theory, it is possible to prove following.
\begin{theorem}
Let $\chi({\bf r},{\bf s})$ a non-negative function with 
support in $\Omega$ and $\rho(r,s|a,b)$ the associated
unnormalized local distribution,
then the inequality
\begin{equation}
\begin{array}{c}
\sum_{a,b} \rho_\Omega^{min}(r_a,s_b|a,b)W(a,b)
\ge \sum_{a,b}\rho(r_a,s_b|a,b)W(a,b)   \\
-\left[2\left( F^{(+)}-F_\Omega^{(-)}\right)\right]^{1/2}.
\end{array}
\end{equation}
holds. 
\end{theorem}
A direct consequence of this theorem and the slackness 
condition~(\ref{second_cond}) for optimality
is the following.
\begin{corollary}
Let $\chi({\bf r},{\bf s})$ a non-negative function with
support in $\Omega$ and $\rho(r,s|a,b)$ 
the associated unnormalized local distribution.
If the inequality
\begin{equation}
\label{cond_zero_chi}
\begin{array}{c}
\sum_{a b}g(r_a,s_b;a,b) \le 
\left\{
R S \beta^2 + \right.
\vspace{1mm} \\
\left.
2\sum_{r s a b}\left[\beta-g(r,s;a,b)\right]\rho(r,s|a,b)
W(a,b)\right\}^{1/2}
\end{array}
\end{equation}
holds with $g(r,s|a,b)=P(r,s|a,b)-\rho(r,s|a,b)$,
then $\chi_\Omega^{min}({\bf r},{\bf s})$ is
equal to zero.
\end{corollary}
Condition~(\ref{cond_zero_chi}) is sufficient for having
$\chi_\Omega^{min}({\bf r},{\bf s})$ equal to zero, but
it is not necessary. A necessary condition can be derived
by computing the lowest eigenvalue of the Hessian of the
objective function $F[\chi]$. Both the necessary and sufficient
conditions allows us to determine the support of the
minimizer $\chi_\Omega^{min}$ once the distribution $\chi$
is enough close to $\chi_\Omega^{min}$.
Thus, the minimization at Step~3 should not stop until
each sequence $({\bf r},{\bf s})$ satisfies the sufficient
condition or does not satisfy the necessary condition,
otherwise the
cleaning up could miss some points where the minimizer 
is equal to zero. However, numerical experiments show 
that the use of these conditions is not necessary
and the number of elements in the sets $\Omega_n$
is always bounded by $d_{NS}+1$, provided that 
Problem~\ref{secondary_problem} is solved by the algorithm
described in the following section.

\subsection{Solving Problem~\ref{secondary_problem}}
\label{sec_algo_probl_2}

There are standard methods for solving Problem~2,
and numerical libraries are available. The interior
point method~\cite{boyd} provides a quadratic
convergence to the solution, meaning that the
number of digits of accuracy is almost doubled 
at each iteration step, once $\chi$ is sufficiently
close to the minimizer. The algorithm uses the
Newton method and needs to solve a set of linear 
equations. Since this can be computationally
demanding in terms of memory, we have implemented 
the solver by using the conjugate gradient method, 
which does not use the Hessian.
Furthermore, if the Hessian
turns out to have a small condition number,
the conjugate gradient method can be much more
efficient than the Newton method, especially
if we do not need to solve Problem~\ref{secondary_problem}
with high accuracy. This is the case in the
initial stage of the computation, when the
set $\Omega$ is growing and does not contain
all the points of the support of $\chi^{min}$.

The conjugate  gradient method iteratively
performs a one-dimensional minimization along
directions that are conjugate with respect to
the Hessian of the objective function~\cite{boyd}. 
The directions are computed iteratively by setting
the first direction equal to the gradient of
the objective function.
The conjugate gradient method is generally used
with unconstrained problems, whereas Problem~2
has the inequality constraints $\chi({\bf r},{\bf s})\ge0$.
To adapt the method to our problem, we 
perform the one-dimensional minimization
in the region where $\chi$ is non-negative.
Whenever an inactive constraint
becomes active or \emph{vice versa}, we 
reset the search direction equal to 
the gradient and restart the generation
of the directions from that point. Once the procedure
terminates, the algorithm provides a list of active
constraints with $\chi_n({\bf r},{\bf s})=0$. Numerical
simulations show that this list is generally complete
and corresponds to the points where the minimizer
$\chi_n^{min}$ is equal to zero. 

In general, the slackness condition~(\ref{slackness1})
is not satisfied by the numerical solution. However,
as previously pointed out,
we can suitably normalize $\chi_n$ so that this condition is
satisfied by $\rho_n(r,s|a,b)$. Thus, 
we will assume that the equality
\begin{equation}
\label{num_slackness}
\sum_{r,s,a,b} \rho_n(r,s|a,b)g_n(r,s;a,b)W(a,b)=0
\end{equation}
holds with $g_n=P-\rho_n$. This also implies that
\begin{equation}
\begin{array}{c}
\alpha_n=\left. \alpha \right|_{\rho=\rho_n}\ge 0  \\
\beta_n=\left. \beta \right|_{\rho=\rho_n}\ge 0.
\end{array}
\end{equation}

\section{Convergence analysis and computational cost}
\label{convergence_sec}
Here, we provide a convergence analysis and we show that
the error on the distance from the local polytope decays
at least as fast as $1/\sqrt{n}$, where $n$ is the number
of iterations. Although the proved convergence is sublinear,
its derivation relies on a very rough estimate of a lower
bound on the optimal value $\chi^{min}$. 
Interestingly, the computed bound on the number of 
required iterations does not depend on the number of 
measurements. Using this bound, we show that the computational 
cost for any given error on the distance grows polynomially 
with the size of the problem input, that is, with $A$, $B$, 
$R$ and $S$, provided that the oracle can be simulated in 
polynomial time.

To prove the convergence, we need to introduce the
dual form of Problem~\ref{main_problem} (See Ref.~\cite{boyd} for 
an introduction to dual theory). The dual form of a minimization 
problem (primal problem) is a maximization problem whose maximum 
is always smaller than or equal to the primal minimum, the 
difference being called \emph{duality gap}. However, if the constraints
of the primal problem satisfy some mild conditions such as Slater's
conditions~\cite{boyd}, then the duality gap is equal to zero.
As previously said, this is the case of Problem~1. 

The dual form is particularly useful for evaluating lower bounds on 
the optimal value of the primal problem. Indeed, the value taken
by the dual objective function in a feasible point of the dual
constraints provides such a bound. After introducing the dual form
of Problem~\ref{main_problem}, we derive the lower bound $F^{(-)}$
on $F^{min}$ given by Eq.~(\ref{lower_bound_0}). Then, we use this 
bound and Eq.~(\ref{sequence_F}) to prove the convergence.

\subsection{Dual problem}
\label{sec_dual}
The dual problem of Problem~\ref{main_problem} is a maximization
problem over the space of values taken by the Lagrange multipliers
$\lambda({\bf r},{\bf s})$ subject to the dual constraints
$\lambda({\bf r},{\bf s})\ge 0$. The dual objective function
is given by the minimum of the Lagrangian $\cal L$, defined by
Eq.~(\ref{lagrangian}), with respect to $\chi$. The dual constraint
is the non-negativity of the Lagrange multipliers, that is,
\begin{equation}
\label{first_dual_constr}
\lambda({\bf r},{\bf s})\ge 0.
\end{equation}
As this minimum
cannot be derived analytically, a standard strategy for getting
an explicit form of the dual objective function is to enlarge the 
space of primal variables and,
correspondly, to increase the number of primal constraints.
The minimum is then evaluated over the enlarged space. In our
case, it is convenient to introduce Eq.~(\ref{local_cond_prob})
and $\rho(r,s|a,b)$ as additional constraints and variables,
respectively. Thus, $F$ is made independent of $\chi$ and
expressed as function of $\rho$. The new optimization problem,
which is equivalent to Problem~\ref{main_problem}, has the
Lagrangian
\begin{eqnarray}
\nonumber
&{\cal L}=F[\rho]-\sum_{{\bf r},{\bf s}}\lambda({\bf r},{\bf s})
\chi({\bf r},{\bf s})+ \sum_{r s a b} W(a,b)\times   & \\
\vspace{1mm}
&\eta(r,s,a,b)\left[\rho(r,s|a,b)-
\sum_{{\bf r},{\bf s}}\delta_{r,r_a}\delta_{s,s_b}\chi({\bf r},{\bf s})
\right],&
\end{eqnarray}
where $\eta(r,s,a,b)$ are the Lagrange multipliers associated with
the added constraints.
To find the minimum of the Lagrangian, we set its derivative 
with respect to the primal variables $\chi$ and $\rho$ equal to zero. 
We get the equations
\begin{eqnarray}
\label{second_dual_constr}
& \sum_{a,b}W(a,b) \eta(r_a,s_b,a,b)=-\lambda({\bf r},{\bf s}) \\
\vspace{1mm}
\label{get_rho}
& \rho(r,s|a,b)=P(r,s|a,b)-\eta(r,s,a,b).
\end{eqnarray}
The first equation does not depend on the primal variables and
sets a constraint on the dual variables. If this constraint
is not satisfied, the dual objective function is equal to
$-\infty$. Thus, its maximum is in the region where 
Eq.~(\ref{second_dual_constr}) is satisfied. Let us add it to the dual
constraint~(\ref{first_dual_constr}). The second stationarity
condition, Eq.~(\ref{get_rho}), gives the optimal $\rho$. By replacing
it in the Lagrangian, we get the dual objective function
\begin{equation}
\begin{array}{r}
F_{dual}=\sum_{r,s,a,b}W(a,b)\eta(r,s,a,b)\times 
\vspace{1mm} \\
\left[P(r,s|a,b)-\frac{\eta(r,s,a,b)}{2}\right].
\end{array}
\end{equation}

Eliminating $\lambda$, which does not appear in the objective
function, the dual constraints~(\ref{first_dual_constr},\ref{second_dual_constr}) 
give the inequality
\begin{equation}
\sum_{a,b}W(a,b) \eta(r_a,s_b;a,b)\le 0.
\end{equation}
Thus, Problem~\ref{main_problem} is equivalent to the following.
\begin{problem}[dual problem of Problem~\ref{main_problem}]
\label{dual_problem}
\begin{equation}
\nonumber
\begin{array}{c}
\max_\eta F_{dual}[\eta] \\
\text{subject to the constraints} \\
\sum_{a,b}W(a,b) \eta(r_a,s_b;a,b)\le 0.
\end{array}
\end{equation}
\end{problem}
The value taken by $F_{dual}$ at a feasible point provides a lower
bound on $F^{min}$. Given any function $\bar\eta(r,s;a,b)$, a feasible
point is 
\begin{equation}
\label{feas_point}
\eta_f(r,s;a,b)\equiv\bar\eta(r,s;a,b)-
\max_{{\bf r},{\bf s}} \sum_{\bar a,\bar b} W(\bar a,\bar b)
\bar\eta(r_{\bar a},s_{\bar b};\bar a,\bar b),
\end{equation}
Indeed
\begin{equation}
\begin{array}{c}
\sum_{a,b}\eta_f(r_a,s_b;a,b)W(a,b)=
\sum_{a,b}\bar\eta(r_a,s_b;a,b) 
\vspace{1mm}  \\
-\max_{{\bf r}',{\bf s}'}
\sum_{a,b}\bar\eta(r_a',s_b';a,b)W(a,b)\le 0.
\end{array}
\end{equation}
The lower bound turns out to be the optimal value $F^{min}$
if the distribution $\rho(r,s|a,b)$ given by Eq.~(\ref{get_rho})
in terms of $\eta=\eta_f$
is solution of the primal Problem~\ref{main_problem}. This
suggests the transformation 
\begin{equation}
\eta_f(r,s;a,b)=P(r,s|a,b)-\rho(r,s|a,b),
\end{equation}
where $\rho(r,s|a,b)$ is some local distribution up to
a normalization constant (in fact, $\rho$ can be any
real function).
Every local distribution induces a lower bound on the
optimal value $F^{min}$. This lower bound turns out to
be an accurate approximation of $F^{min}$ if $\rho$ is
close enough to the optimal local distribution. Using the
last equation and Eq.~(\ref{feas_point}), we get the 
lower bound~(\ref{lower_bound_0}) from $F_{dual}$.

The dual problem of Problem~2 is similar to Problem~\ref{dual_problem},
but the constraints have to hold for sequences $({\bf r},{\bf s})$ in 
$\Omega$.
\begin{problem}[dual problem of Problem~\ref{secondary_problem}]
\begin{equation}
\nonumber
\begin{array}{c}
\max_\eta F_{dual}[\eta] \\
\text{subject to the constraints} \\
({\bf r},{\bf s})\in\Omega\Rightarrow \sum_{a,b}W(a,b) \eta(r_a,s_b;a,b)\le 0.
\end{array}
\end{equation}
\end{problem}
This dual problem induces the lower bound $F_\Omega^{min}$ on the optimal value
of Problem~\ref{secondary_problem} [Eq.~(\ref{lower_bound_Omega})].

\subsection{Convergence and polynomial cost}
\label{sec_conve}
Let $\rho_n(r,s|a,b)$ be the local distribution computed
at Step~3 of Algorithm~\ref{main_algo}. From the lower
bound~(\ref{lower_bound_0}), we have 
\begin{equation}
\label{lower_bound}
\begin{array}{c}
F^{min}\ge F_n-\frac{R S}{2}\alpha_n^2+   
\vspace{1mm}  \\
\sum_{r,s,a,b} W(a,b)
\rho_n(r,s|a,b) \left[g_n(r,s;a,b)-\alpha_n\right],
\end{array}
\end{equation}
where $\alpha_n$ is given by Eq.~(\ref{alpha_n}) and
$g_n=P-\rho_n$. 
The last term is equal to zero because of Eq.~(\ref{num_slackness})
The second term at the right-hand side of Ineq.~(\ref{lower_bound})
is bound from below by $-\alpha_n[1+(R S)^{1/2}]$ ($\alpha_n$ is
positive). This can be shown by minimizing it under the constraint 
that the last term is equal to zero.
Since $\left. F_{dual}\right|_{\eta=\eta_f}$ is a lower bound
of $F^{min}$, we have that
\begin{equation}
\label{dual_bound}
F^{min}\ge  F_n-\frac{R S}{2}\alpha_n^2 -[1+(R S)^{1/2}]\alpha_n.
\end{equation}
As $\alpha_n$ is not greater than $1$, the factor $\alpha_n^2$
in the right-hand side of the inequality can be replaced
by $\alpha_n$,  so that we have
\begin{equation}
\alpha_n \ge 2\frac{F_n-F^{min}}{R S+2+2(R S)^{1/2}},
\end{equation}
which gives with Ineq.~(\ref{num_monotone}) the following
\begin{equation}
F_{n}^{min}-F_{n+1}^{min}\ge 2 (1-\gamma)
\left(\frac{F_n-F^{min}}{R S+2+2(R S)^{1/2}}\right)^2.
\end{equation}
Summing over $n=0,\dots,\infty$ both terms of the inequality,
we have
\begin{equation}
\label{series_bound}
F_{0}^{min}-F^{min}_\infty\ge 2 (1-\gamma)\sum_{k=0}^\infty
\left(\frac{F_k-F^{min}}{R S+2+2(R S)^{1/2}}\right)^2.
\end{equation}
The inequality implies that
the series at the right-hand side converges to a finite number and,
thus,
\begin{equation}
\lim_{n\rightarrow\infty} F_n= F^{min}.
\end{equation}
In particular, the terms in the series go to zero at least as
fast as $1/n$. Indeed, since $F_n-F^{min}$ is a decreasing
sequence, we have from Eq.~(\ref{series_bound}) 
\begin{eqnarray}
\label{bound_err}
F_0^{min}-F^{min}_\infty&\ge& 2 (1-\gamma)\sum_{k=0}^n
\left(\frac{F_k-F^{min}}{R S+2+2(R S)^{1/2}}\right)^2 \\
\nonumber
&\ge&
2 (n+1)(1-\gamma)\left(\frac{F_n-F^{min}}{R S+2+2(R S)^{1/2}}\right)^2.
\end{eqnarray}
The quantity at the left-hand side does not depend on $n$
and is not greater than $1/2$. Indeed, using the identity
\begin{equation}
\begin{array}{c}
\sum_{r,s,a,b}W(a,b) \rho_0^{min}(r,s|a,b)\times  
\vspace{1mm} \\
\left[P(r,s|a,b)-\rho_0^{min}(r,s|a,b)\right]=0,
\end{array}
\end{equation}
we have
\begin{equation}
\begin{array}{l}
F_0^{min}-F^{min}_\infty\le F_0^{min}= \\
\sum_{r,s,a,b} W(a,b)
\frac{\left[P(r,s|a,b)-\rho_0^{min}(r,s|a,b)\right]^2}{2} 
\vspace{1mm} \\
= \sum_{r,s,a,b}W(a,b) \frac{P^2(r,s|a,b)-(\rho_0^{\min})^2(r,s|a,b)}{2} 
\vspace{1mm} \\ 
\le \sum_{r,s,a,b}W(a,b) \frac{P^2(r,s|a,b)}{2} \le \frac{1}{2}.
\end{array}
\end{equation}
The last inequality and Ineq.~(\ref{bound_err}) give a bound
on the accuracy reached at step $n$,
\begin{equation}
\label{accuracy}
F_n-F^{min}\le \frac{R S+2+2(R S)^{1/2}}{2 \sqrt{(1-\gamma)(n+1)}}.
\end{equation}
Thus, the error decreases at least as fast as
$1/\sqrt{n}$. Although the convergence is sublinear, we derived
this inequality by using Ineq.~(\ref{dual_bound}), which provides
a quite loose bound on the optimal value $\chi^{min}$. Nonetheless,
the constraint set by Ineq.~(\ref{accuracy}) on the accuracy is
strong enough to imply the polynomial convergence of the
algorithm, provided that the oracle can be simulated in
polynomial time. Indeed, the inequality implies that the
number of steps required to reach a given accuracy does not
grow faster than $(R S)^2$. Since the computational cost
of completing each step is polynomial, the overall algorithm
has polynomial cost. More precisely, each step is completed by
solving a quadratic minimization problem. If we do not rely on
specific structure of the quadratic problem, its computational
cost does not grow faster than $max\{n_1^3,n_1^2 n_2,D\}$~\cite{boyd},
where $n_1$, $n_2$ and $D$ are the number of variables,
the number of constraints and the cost of evaluating first
and second derivatives of the objective and constraint
functions. The numbers $n_1$ and $n_2$ are equal and 
$D$ is equal to $n_1^2 (A+B)$. As the number of vertices
in the set $\Omega$ is not greater than the number of
iterations $n$, if $A+B>n$, then
the computational cost at each iteration 
does not grow faster than $n^2 (A+B)$. This implies
that the computational cost of Algorithm~\ref{main_algo}
scales at most as
\begin{equation}
\frac{(R S)^{3} (A+B)}{\epsilon^6},
\end{equation}
$\epsilon$ being the accuracy. Thus, the bound 
is linear in the number of measurements. This
bound holds asymptotically for a sufficiently high number
of measurements, given a fixed error. If $A+B$ is smaller
than the number of iterations, then the running time of 
the overall algorithm scales at most as
\begin{equation}
(A B)^3(R S)^5,
\end{equation}
provided that the number of vertices is not greater
than $d_{NS}$, which is guaranteed by the cleaning
up procedure.
Numerical tests are in good agreement with this estimate,
the linear bound not being saturated for the considered
accuracy and number of measurements.

\subsection{Simulation of the oracle}
\label{sec_oracle}
We have shown that the cost of computing the distance from
the local polytope grows polynomially provided that we
have access to the oracle. But what is the computational
complexity of the oracle? In the case of measurements with
two outcomes, we have seen that the solution of the oracle 
is equivalent to finding the minimal energy of a particular
class of Ising spin glasses. These problems are known 
to be hard to 
solve. However, the oracle has a particular structure 
that can enormously reduce the complexity of the 
problem and make it numerically tractable in most of the
cases or, possibly, in all the cases. Indeed, the
couplings of the Ising spin model are constrained
by the nonsignaling conditions on $P(r,s|a,b)$
and the optimality
conditions~(\ref{first_cond}-\ref{third_cond}). Furthermore, 
Hamiltonian~(\ref{spin_glass}) is characterized by two classes 
of spins, described by the variables $r_k$ and $s_k$
respectively, and each element in one class is coupled 
only to elements in the other class. This particular 
structure suggests the following block-maximization algorithm 
for solving the oracle.
\begin{algorithm}
\label{oracle_algo}
Input: $g(r,s;a,b)$
\begin{enumerate}
\item 
\label{step1}
Generate a random sequence $\bf r$.
\item 
\label{step2}
Maximize $\sum_{a,b}g(r_a,s_b;a,b) W(a,b)$ with respect
to the sequence $\bf s$ (see later discussion).
\item Maximize $\sum_{a,b}g(r_a,s_b;a,b) W(a,b)$ with respect
to the sequence $\bf r$.
\item Repeat from Step~\ref{step2} until the block-maximizations
stop making progress.
\end{enumerate}
\end{algorithm}
Numerical tests show that this algorithm, when it is used 
for computing the distance from the local polytope, stops
after few iterations. Furthermore, only few trials of
the initial random sequence $\bf r$ are required for
a convergence of Algorithm~\ref{main_algo}. We also
note that the probability is a successful simulation
of the oracle increases when $\chi$ is close to the optimal 
solution $\chi^{min}$, suggesting that the optimality
conditions~(\ref{first_cond}-\ref{third_cond}) play
a pivotal role on the computational complexity of the
oracle. Pragmatically,
we have chosen the number of trials equal to $d_{NS}$ so that
the computational cost of simulating the oracle contributes
to the overall running time with a constant multiplicative 
factor and, thus, the sixth-power law of the oracle-assisted
algorithm is not affected.

Before discussing the numerical results, let us explain how
the maximization on blocks is performed. Let us consider the 
maximization with respect to $\bf r$, as the optimization with 
respect to $\bf s$ has an identical procedure. We have
\begin{equation}
\begin{array}{c}
\max_{\bf r}\sum_{a,b}W(a,b) g(r_a,s_b;a,b)=
\vspace{1mm} \\
\sum_a \max_r \sum_b g(r,s_b;a,b) W(a,b)\equiv
\vspace{1mm} \\
\sum_a \max_r \tilde g(r,{\bf s};a).
\end{array}
\end{equation}
Thus, the maximum is found by maximizing the function
$\tilde g(r,{\bf s};a)$ with respect to the discrete
variable $r$ for every $a$. Taking into account the sum over
$b$ required for generating $\tilde g$, the computational 
cost of the block-maximization is proportional to $R A B$.
Thus, it does not grow more than linearly with respect to the 
size of the problem input, that is, $R S A B$.

\section{Numerical tests}
\label{sec_numerical}
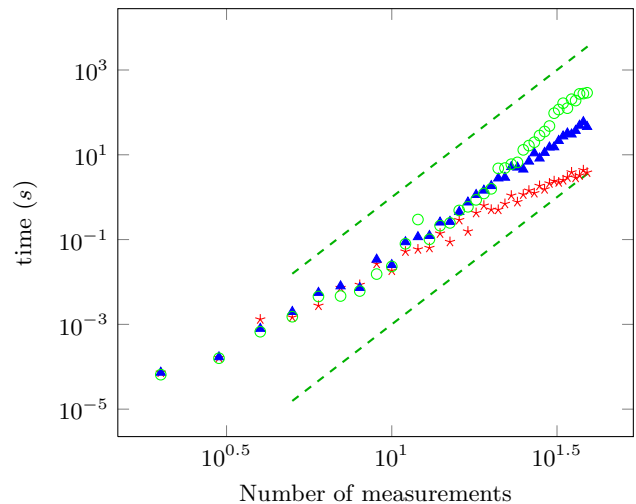
\begin{figure}
\begin{tikzpicture}
\begin{loglogaxis}
[ylabel=time ($s$),
xlabel={Number of measurements},
]
\addplot[red, only marks, mark=star] table[x index={0}, y index={1}]{times_1e-3.txt};
\addplot[blue, only marks, mark=triangle*] table[x index={0}, y index={1}]{times_1e-4.txt};
\addplot[green, only marks, mark=o] table[x index={0}, y index={1}]{times_1e-5.txt};
\addplot[green!70!black,domain=5:40, samples=200, dashed, thick]{10^(-6)*x^6};
\addplot[green!70!black,domain=5:40, samples=200, dashed, thick]{10^(-9)*x^6};
\end{loglogaxis}
\end{tikzpicture}
\caption{Time required for computing the distance from the local polytope
as a function of the number of measurements (log-log scale) with accuracy equal to
$10^{-3}$, $10^{-4}$ and $10^{-5}$ (red, blue and green points, respectively).}
\label{loc_pol_dist}
\end{figure}

In the previous sections, we have introduced an algorithm 
that computes the distance from the local polytope in polynomial
time, provided that we have access to the oracle. 
Surprisingly, in every simulation performed
on entangled qubits, the algorithm implementing the oracle
successfully finds the solution in polynomial time. More
precisely, the algorithm finds a sequence $({\bf r},{\bf s})$
sufficiently close to the maximum to guarantee a convergence
of Algorithm~\ref{main_algo} to the solution of Problem~1.
Interestingly, the probability of a successful simulation
of the oracle increases as $\chi$ approaches the solution.
This suggests that the optimality conditions~(\ref{first_cond}-\ref{third_cond})
play a fundamental role on the computational 
complexity of the oracle. To check that the algorithm successfully
finds the optimizer $\chi^{min}$ up to the desired accuracy,
we have solved the oracle with a brute force search at the
end of the computation whenever
this was possible in a reasonable time. All the checks show
that the solution is found within the desired accuracy.

In the tests, we have considered the case of maximally
entangled states, Werner states and pure non-maximally entangled
states. We always observe a power law of the running time
in accordance with the theoretical analysis given in
Sec.~\ref{sec_conve}. Let us discuss the case of entangled
qubits in a pure quantum state.

\subsection{Maximally entangled state}

In Fig.~\ref{loc_pol_dist}, we report
the time required for computing the distance from the
local polytope as a function of the number of measurements, 
$M$, in log-log scale. The distance has been evaluated with 
accuracy equal to $10^{-3}$, $10^{-4}$ and $10^{-5}$
(red, blue and green points, respectively). We have
considered the case of planar measurements on the
Bloch sphere.
For the sake of comparison, we have also
plotted the functions $10^{-6} M^6$ and
$10^{-9} M^6$ (dashed lines). The data 
are compatible with the theoretical power law
derived previously. 
Other simulations have been performed with random 
measurements and we always observed the same power 
law. For a number of measurements below $28$, we
have solved the oracle with a brute force search
at the end of the computation and we have always 
found that Algorithm~\ref{main_algo} successfully
converged to the solution within the desired
accuracy.

\subsection{Non-maximally entangled state}

In the case of the non-maximally entangled state 
\begin{equation}
|\psi\rangle=\frac{|0 0\rangle+\gamma|1 1\rangle}{\sqrt{1+\gamma^2}},
\end{equation}
with $\gamma\in[0,1]$, we have considered
planar measurements orthogonal to the
Bloch vector ${\vec v}_z\equiv(0,0,1)$ so that the
marginal distributions are unbiased, as well
as planar measurements lying in the plane
containing ${\vec v}_z$ (biased marginal distributions).

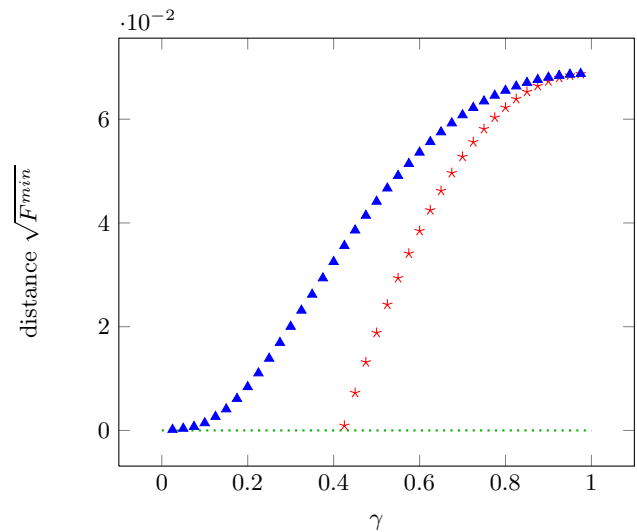
\begin{figure}
\begin{tikzpicture}
\begin{axis}
[ylabel=distance $\sqrt{F^{min}}$,
xlabel={$\gamma$},
]
\addplot[red, only marks, mark=star] table[x index={0}, y index={1}]{dist_unbiased.txt};
\addplot[blue, only marks, mark=triangle*] table[x index={0}, y index={1}]{dist_biased.txt};
\addplot[green!70!black,domain=0:1, samples=200, dotted, thick]{0.};
\end{axis}
\end{tikzpicture}
\caption{Distance from the local polytope as a function of $\gamma$ in the
unbiased (red stars) and biased case (blue triangles).}
\label{fig_dist}
\end{figure}

In Fig.~\ref{fig_dist}, we report the distance
from the local polytope as a function of $\gamma$
with $10$ measurements. The distance changes slightly
for higher numbers of measurements.
In the unbiased case, the distance goes to zero
for $\gamma$ equal to about $0.4$, whereas
the correlations become local for $\gamma=0$
in the biased case.

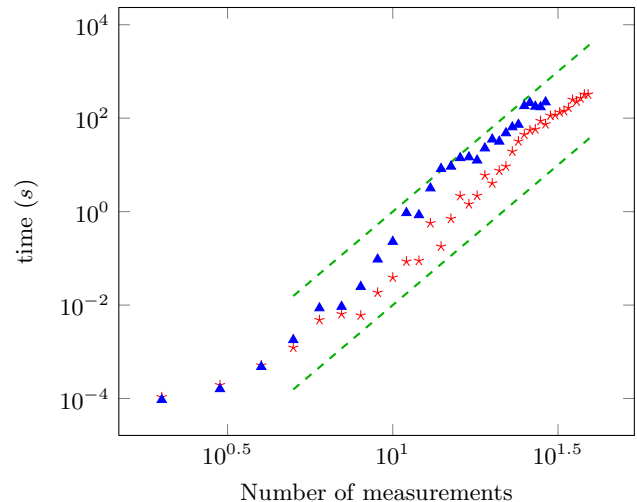
\begin{figure}
\begin{tikzpicture}
\begin{loglogaxis}
[ylabel=time ($s$),
xlabel={Number of measurements},
]
\addplot[red, only marks, mark=star] table[x index={0}, y index={1}]{times_g=0.8_unbiased.txt};
\addplot[blue, only marks, mark=triangle*] table[x index={0}, y index={1}]{times_g=0.6_unbiased.txt};
\addplot[green!70!black,domain=5:40, samples=200, dashed, thick]{10^(-6)*x^6};
\addplot[green!70!black,domain=5:40, samples=200, dashed, thick]{10^(-8)*x^6};
\end{loglogaxis}
\end{tikzpicture}
\caption{Time required for computing the distance from the local polytope
as a function of the number of measurements (log-log scale) in the unbiased case
for $\gamma=0.8$ (red stars) and $\gamma=0.6$ (blue triangles). 
The green lines are the functions $10^{-6} M^6$ and $10^{-8} M^6$.
The accuracy is $10^{-5}$.}
\label{fig_unbiased}
\end{figure}

\begin{figure}
\begin{tikzpicture}
\begin{loglogaxis}
[ylabel=time ($s$),
xlabel={Number of measurements},
]
\addplot[red, only marks, mark=star] table[x index={0}, y index={1}]{times_g=0.8_biased.txt};
\addplot[blue, only marks, mark=triangle*] table[x index={0}, y index={1}]{times_g=0.6_biased.txt};
\addplot[green, only marks, mark=o] table[x index={0}, y index={1}]{times_g=0.4_biased.txt};
\addplot[green!70!black,domain=5:40, samples=200, dashed, thick]{10^(-6)*x^6};
\addplot[green!70!black,domain=5:40, samples=200, dashed, thick]{10^(-8)*x^6};
\end{loglogaxis}
\end{tikzpicture}
\caption{The same as Fig.~\ref{fig_biased} in the biased case 
for $\gamma=0.8$ (red stars), $\gamma=0.6$ (blue triangles) and $\gamma=0.4$ (green circles).}
\label{fig_biased}
\end{figure}
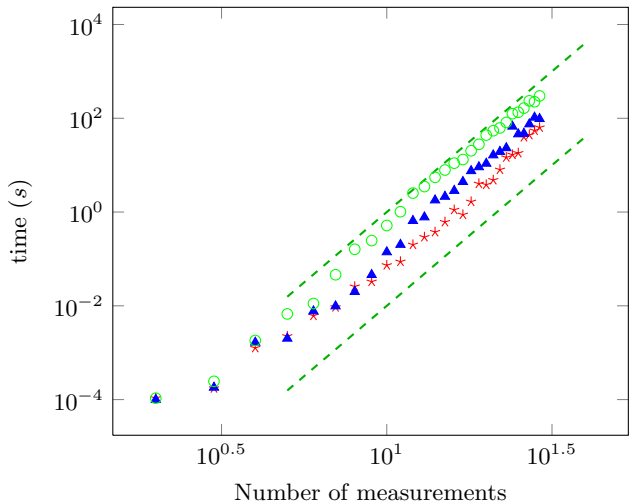

In Figs.~\ref{fig_unbiased},\ref{fig_biased}, the running time as a function
of the number of measurements is reported for the biased and unbiased cases,
respectively. The power law is again in accordance with the theoretical
analysis. As done for the maximally entangled case, we have checked 
the convergence to the solution by solving the
oracle with a brute force search for a number of measurements up to $28$.

\section{Conclusion}
In conclusion, we have presented an algorithm that computes the distance
of a given non-signaling correlation to
the local polytope. The running time with given
arbitrary accuracy scales polynomially, granted the access to a 
oracle determining the optimal locality bound of a Bell inequality.
We also propose an algorithm for simulating the oracle.
In all the numerical tests, the overall algorithm successfully computes
the distance with the desired accuracy and a scaling of the 
running time in agreement with the bound 
theoretically derived for the oracle-assisted algorithm. These
results take us to question whether the non-locality problem is actually
computationally hard, as generally believed because of a result of
Pitowski~\cite{pitowski}. Our algorithm opens the way to tackle many
unsolved problems in quantum theory, such as the nonlocality of
Werner states. Because of the relation between the non-locality problem 
and computational complexity theory, the latter can profit from a deeper
understanding of the complexity of the former. This study would be
particularly interesting if deciding the membership
to the local polytope turned out to be a NP-complete problem, as suggested
by Pitowski's work. In this
case, our work and its further refinements could provide alternative
algorithms to solve some instances of computationally hard problems.
Thus, a fundamental task of future investigations will be to determine the 
complexity class of the oracle.

{\it Acknowledgments.} 
We wish to thank Arne Hansen for valuable comments and suggestions.
This work is supported by the Swiss National Science Foundation, the NCCR QSIT, 
and the Hasler foundation through the project "Information-Theoretic Analysis of 
Experimental Qudit Correlations".

\bibliography{biblio.bib}

\end{document}